\documentclass[9pt,twocolumn,twoside]{osajnl}

\usepackage{amsmath}
\usepackage{slashed}

\journal{josaa} 

\setboolean{shortarticle}{false}

\title{Ray-Wave Duality of Electromagnetic Fields: A Feynman Path integral Approach to Classical Vectorial Imaging}

\author[1,*]{James Babington}

\affil[1]{Thales Optronics Ltd, Linthouse Road, Glasgow, G51 4BZ, UK.}

\affil[*]{Corresponding author: james.babington@uk.thalesgroup.com}

\ociscodes{(080.1510)   Propagation methods;  (080.5084)   Phase space methods of analysis ; (110.2760)   Gradient-index lenses; (000.6800)   Theoretical physics; (050.5082)   Phase space in wave optics.}

\begin{abstract}
We consider how vectorial aspects (polarization) of light propagation can be implemented, and its origin, within a Feynman path integral approach. A key part of this scheme is in generalising the standard optical path length integral from a scalar to a matrix quantity. Reparametrization invariance along the rays allows a covariant formulation where propagation can take place along a general curve. A general gradient index background is used to demonstrate the scheme. This affords a description of classical imaging optics when the polarization aspects may be varying rapidly and cannot be neglected.
\end{abstract}

\setboolean{displaycopyright}{true}

\begin{document}
\maketitle
\section{Introduction}

The connections between classical optics, quantum mechanics and general relativity are perhaps not surprising given their shared historical heritages~\cite{misner1973gravitation}. One aspect of these relationships is ray-wave duality. By this we mean the following: It should be possible to calculate any physical quantity of interest by either using a ray description \emph{or} a wave description. In particular, classical optics has a ray-wave duality in just the same manner as present in quantum mechanics (in terms of photons and electrons, quantum mechanics was developed to accommodate the ray-wave duality associated with classical light and the photon). 

Ray-wave dualities have been previously considered and used from a variety of different standpoints. In~\cite{Gloge:69}, wave optics and ray optics are shown to have the same relation as classical mechanics has to quantum mechanics (of a point particle). The thrust of this work is in showing the same structure of the two formalisms. In particular, the relation between the Helmholtz equation and its slow envelope approximation is shown to have an analogous structure to the relativistic Klein Gordon equation and the Schr\"{o}dinger equation when making a non-relativistic approximation. Differently in~\cite{forbes98}, the connections between rays and waves are made by a blurring of curves in phase space. In~ \cite{testorf2001}, suggestions are made for how to interface ray tracing methods with wave methods to calculate diffraction, whilst in~\cite{Gitin:13}, the eikonal is used to make the connection between ray and wave optics. Finally in~\cite{Mout2016}, Monte Carlo ray tracing is performed in conjunction with a Huygens-Fresnel path integration method to evaluate field propagation and point spread functions (these would normally be given by a wave optics calculation). All of the approaches, past and present, show the rich structures encountered between dual descriptions of the same underlying optics.

In a recent paper~\cite{Babington:18}, I considered the connection between rays and waves in classical optics and how this duality is made manifest with the Feynman path integral framework. In particular, the proper implementation of constraint equations is central to this scheme. The notion of a gauge symmetry was introduced as a means to make the optical path length look more like a Lagrangian one would usually encounter in classical mechanics. An important aspect of this approach is the removal of gauge degrees of freedom to obtain physical results. An interesting practical application has been given of this approach in~\cite{Wan:20} and~\cite{Wang:21} (see also \cite{Shen:20} and \cite{PhysRevA.102.031501}) in terms of digitally structuring light and the use of $SU(2)$ group theory states. Actually, the path integral is an ideal tool for studying $SU(2)$ as it is afforded a compact group representation and the path integral becomes tractable.

In this paper we look to further extend this work by firstly elucidating more on gauge invariance and fixing in path integrals, and secondly, moving to a full \emph{vector} description (thereby including polarization). Details are presented on gauge invariance and gauge fixing in the path integral in terms of the Faddeev-Popov determinant~\cite{chaichian2001}. This procedure should not be confused with the gauge transformations associated with writing the electric and magnetic fields in terms of a scalar and vector potential. The question of incorporating vectorial aspects into path integrals has been considered in~\cite{Dimant:10}, where spin coherent states are used to include polarization, and in~\cite{Gersten:87}, where a first order matrix form of Maxwell's equations are employed. The vector nature of the electromagnetic field in the current work is incorporated into the path integral formalism ab initio, but requires us to move away from the simple scalar valued optical path length function. It is required to move to a second rank tensor optical path length \emph{matrix}. This allows initial and final conditions that involve both spatial positions and states of polarization. 

The two examples we consider are based on the quadratic GRIN media, as this provides a useful theoretical laboratory for developing intuition for path integral methods. It is also a topic of current practical value and research~\cite{ocier2020}. The first example is a more detailed treatment of the scalar field version, where in particular one can calculate explicitly the Gouy phase. The second example is an evaluation of the longitudinal component of field propagation that is strikingly similar to the Gouy phase. Additionally we use material data published in~\cite{ocier2020} to calculate parameters associated with our GRIN examples.

\section{Path Integral Representations for scalar fields}
\label{sec:review}

\begin{figure}
\begin{center}
\includegraphics[scale=0.45]{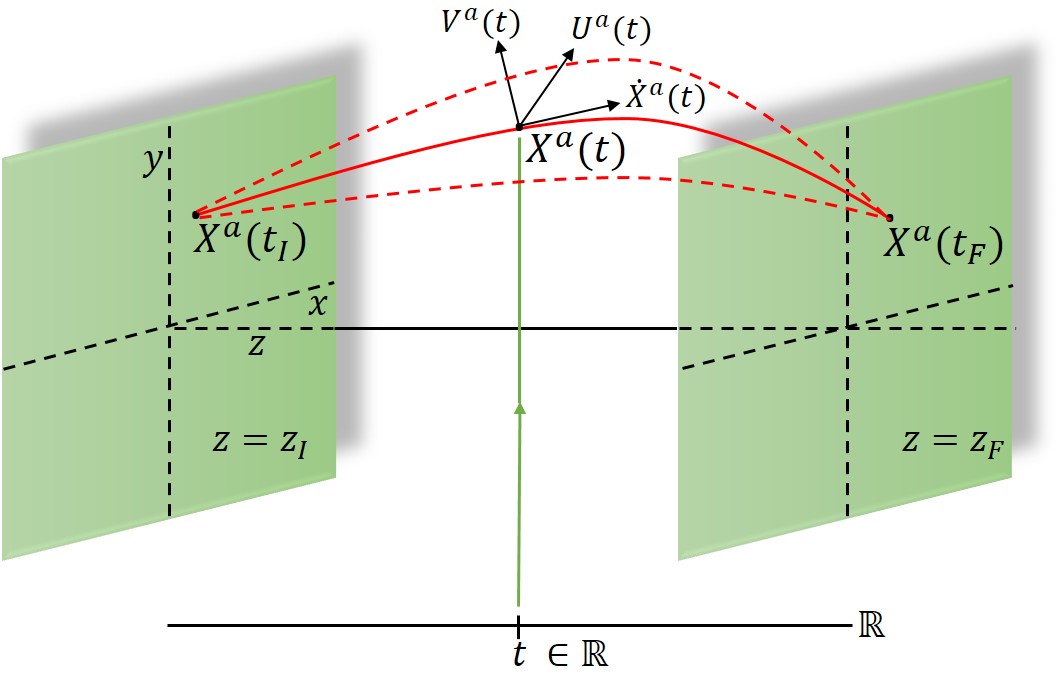} 
\caption{Basic construction and notation used for the path integral description. $\dot{X}^a(t)$ is the tangent vector to the curve (red line) followed by an optical ray in a background refractive index $n(X^a)$. $V^a(t)$ and $U^a(t)$ are orthogonal vectors that represent directions of transverse polarisation.} \label{fig:setup}
\end{center} 
\end{figure}

Firstly, let us recap~\cite{Babington:18} the path integral form that we will apply to a general refractive index media of the gradient index (GRIN) type. We choose a GRIN media because it brings out general features that may be overlooked in a piecewise constant refractive index background. The standard optical path length (OP) functional~\footnote{Also called the action.} used in classical optics is given by
\begin{eqnarray}
S[X,\dot{X}]&=&\int^{t_F}_{t_I} dt \sqrt{ n^2\delta_{ab}\dot{X}^a\dot{X}^b }.\label{eq:opl}
\end{eqnarray}
In the above $a,b = 1, 2, 3,$ are spatial indices (e.g $x=X^1, y=X^2, z=X^3$) and $\delta_{ab}$ is the three dimensional Kronecker delta. The rays propagate in a spatially varying background refractive index $n=n(X^a)$. In addition, repeated indices are summed over as per the Einstein summation convention~\cite{misner1973gravitation}. The $(X^a(t),\dot{X}^a(t))$ are the \textit{(position coordinates, velocities)} on a curve, parametrised by the real parameter $t$ (where the dot denotes a $t$-derivative). The curves the light ray follow (with $\dot{X}^a(t)$ tangent to the curve at any point) are indicated in Figure~\ref{fig:setup} by the solid red line (with neighbouring rays shown by broken lines). The square root in Equation~(\ref{eq:opl}) can be removed by introducing an einbein $\Lambda(t)$, which is a non-dynamical Lagrange multiplier
\begin{eqnarray}
S[X,\dot{X}, \Lambda]&=&\frac{1}{2\lambda}\int^{t_F}_{t_I} dt \frac{1}{\Lambda(t)}\delta_{ab}\dot{X}^a\dot{X}^b +\Lambda(t)\lambda^2 n^2.\label{eq:opl2}
\end{eqnarray}
where $\lambda$ is the wavelength of light. Since $X^a$ has the dimensions of length and the parameter $t$ is dimensionless, we require the introduction of $\lambda$ to keep consistent dimensions. By using the equation of motion for $\Lambda(t)$, this can be substituted into Equation~(\ref{eq:opl2}) thereby eliminating $\Lambda(t)$ and recovering Equation~(\ref{eq:opl}). The use of the Lagrange multipliers is exactly the program followed by Dirac in quantizing a system with redundant degrees of freedom (constraints), albeit now applied to the path integral version. In terms of the phase space variables $(X^a,P_b)$ where $P_b$ is the conjugate momentum to $X^a$, the OP functional is
\begin{eqnarray}
S[X,P,\Lambda] & = & \int^{t_F}_{t_I} dt P_a\dot{X}^a -H(X,P,\Lambda), \label{eq:action} \\
H(X,P,\Lambda)&\equiv& \lambda\Lambda(t)( \delta^{ab}P_aP_b - n^2)/2. \label{eq:hamiltonian}
\end{eqnarray}
We have introduced the Hamliltonian $H(X,P,\Lambda)$ that is a function of auxillary field $\Lambda(t)$. Note also that the momenta are dimensionless. Upon elimination of the momenta $P_a$ via the equations of motion, one recovers Equation~(\ref{eq:opl2}). We have three forms for the OP functional (nonlinear, second-order, and first-order Hamiltonian) that are all equivalent.

Note that in all forms, the action  is reparametrization invariant, where $t \rightarrow f(t)$ is a symmetry. For an infinitesimal reparametrization transformation $t \rightarrow t + \epsilon $ we have.
\begin{eqnarray}
\delta t &=& \epsilon, \\
\delta X^a &=& \epsilon \dot{X}^a, \\
\delta P_a &=& \epsilon \dot{P}_a, \\
\delta \Lambda &=& \frac{d (\epsilon \Lambda)}{dt}. \label{eq:einbeintrans}
\end{eqnarray}
These play an important role in reducing the number of degrees of freedom in the path integral summation. The equation of motion for the auxillary field $\Lambda (t)$ of Equation~(\ref{eq:action}) is just
\begin{eqnarray}
\frac{\delta S[X,P,\Lambda]}{\delta\Lambda(t)} & = & \lambda( \delta^{ab}P_aP_b - n^2)/2 = 0, \\
&\Rightarrow & \nonumber \\
 H(X,P,\Lambda)&=&0. 
\end{eqnarray}
So the Hamiltonian vanishes (as it should) because it is a constraint equation for the rays. With the standard quantum mechanical operator replacement $P_a\rightarrow i\partial_a$ (see~\cite{Gloge:69}) one obtains the the Helmholtz equation when the momentum operators are represented on spatial functions $\phi(X)$, viz  
\begin{equation}
\left(\lambda^2\delta^{ab}\partial_{a}\partial_{b} +n^2\right)\phi(X)=0. \label{eq:helmholtz}
\end{equation}
However, the Hamiltonian also generates evolution in the evolution parameter $t$. The key observation here is that evolution in $t$ is pure gauge. In the Hamiltonian formalism, the generator of a gauge transformation vanishes weakly. We represent this here by lifting the field $\phi (X)\rightarrow \Phi(X,t)$ that incorporates the evolution parameter $t$. Then
\begin{eqnarray}
\Lambda(t)\left(\lambda^2\delta^{ab}\partial_{a}\partial_{b} +n^2\right)\Phi(X,t)= i\frac{\partial \Phi(X,t)}{\partial t}, \\
\hat{H}(X,i\partial, \Lambda)\Phi(X,t)= i\lambda\frac{\partial \Phi(X,t)}{\partial t}.
\end{eqnarray}
One now has a Schr\"{o}dinger type equation but without having singled out the $z$-direction as the parameter and direction of propagation, thereby maintaining covariance~\cite{misner1973gravitation}. Note also here the connection to the heat kernel approach in field theory. This has a standard path integral description in phase space~\cite{chaichian2001} given by
\begin{eqnarray}
\langle X_{F},t_{F} \vert X_{I},t_{I} \rangle_{\Lambda} &=& \int[dX(t)dP(t)]_{ _I}^{_F}e^{ \frac{i}{\lambda} S[X,P,\Lambda]},  \label{eq:scalarpi} \\
&=& \int[dX(t)]_{ _I}^{_F}e^{ \frac{i}{\lambda} S[X,\dot{X},\Lambda]}.
\end{eqnarray}
The first path integral uses exactly the Hamiltonian (Equation~(\ref{eq:hamiltonian})) we constructed earlier in the ray formulation. The propagation amplitude $\langle X_{F},t_{F} \vert X_{I},t_{I} \rangle_{\Lambda}$ at this point depends on the auxillary field; it remains to eliminate it in some manner to make the amplitude physical. Integration over the einbein field $\Lambda(t)$ produces 
\begin{eqnarray}
\langle X_{F},t_{F} \vert X_{I},t_{I} \rangle &=& \int[dX(t)dP(t)\Lambda(t)]_{ _I}^{_F}e^{ \frac{i}{\lambda} S[X,P,\Lambda]}, \\  \label{eq:scalarpi2}
&=& \int[dX(t)dP(t)]_{ _I}^{_F}e^{ \frac{i}{\lambda}\int^{t_F}_{t_I} dt P_a\dot{X}^a } \nonumber \\\times \delta (\delta^{ab}P_aP_b - n^2). \nonumber \\
\end{eqnarray}
The delta functional $\delta (\delta^{ab}P_aP_b - n^2)$ is the way in which the Helmholtz equation manifests itself as a constraint equation. Thus the Hamiltonian itself vanishes if we use the equation of motion for the einbein field. This is because the Hamiltonian generates evolution in the parameter $t$, but any change in $t$ is pure gauge!

Since the OP functional is reparametrization invariant, the path integral Equation~(\ref{eq:scalarpi}) is actually integrating over an infinite number of copies of the same action. This leads to a gauge volume, $Vol(gauge):=\mathcal{G}$, that is an over counting over all the possible paths used in the integration. Fortuitously, $\mathcal{G} $ can be factored out as an overall multiplicative factor from the integration by the inclusion of a gauge fixing functional term, $\mathcal{F}(\Lambda)= 0$. By including in the path integral a delta functional $\delta[ \mathcal{F}(\Lambda)]$, we can eliminate all the identical copies being summed over~\cite{chaichian2001}, whilst allowing the curves to be physically realized. However, the inclusion of the delta functional into the path integral must be accompanied by a Faddeev-Popov determinant, $\Delta_{FP}$, to ensure a correct integration measure~\cite{Mottola95}. This factor is a necessary inclusion because it \emph{removes} degrees of freedom from the path integral that are unphysical. By integrating over the gauge group, the delta functional gives rise to $\Delta_{FP}$ in the following way
\begin{eqnarray}
\mathbb{1} &=& \Delta_{FP}\int [d \epsilon] \delta[\mathcal{F}(\Lambda(\epsilon)], \label{eq:fpdet} \\
\mathcal{G} &=& \int [d \epsilon].
\end{eqnarray}
We have used the symbol '$\mathbb{1}$' to indicate the resolution of unity, i.e. the number $1$ but used for insertion into integrals, much in the same way as occurs for eigen-functions. In its most usual usage, this determinant is represented in terms of Grassmanian valued fields~\cite{chaichian2001}. The effect of this normally is to maintain covariance in perturbative calculations whilst allowing the so called 'ghost fields' to cancel the extra unphysical degrees of freedom. In our case, however, this is not necessary. The key property we require is just a reduction from three to two degrees of freedom at the expense of covariance. The form of $\Delta_{FP}$ can be found from the gauge transformation given in Equation~(\ref{eq:einbeintrans}). To be definite and as an example, we choose the gauge $\mathcal{F}(\Lambda) =\Lambda -1 = 0$. Then it follows that
\begin{eqnarray}
\Delta_{FP} &=& \det \left[\frac{\delta \mathcal{F}}{\delta \epsilon}\right]_{\mathcal{F}=0}, \\
&=& \det \left[\frac{d ( \Lambda)}{dt}\right]_{\Lambda = 1 }, \\
&=& \det \left[ \frac{d}{dt} \right] . \label{eq:fpdet2}
\end{eqnarray}
Thus we have a representation of the Faddeev-Popov determinant that can be used directly to remove the unphysical degrees of freedom from path integral. Inserting Equation~(\ref{eq:fpdet}) directly into the path integral Equation~(\ref{eq:scalarpi2}) gives
\begin{eqnarray}
\langle X_{F},t_{F} \vert X_{I},t_{I} \rangle = \int^F_I [dX(t)dP(t)d\Lambda (t)]_{ _I}^{_F}\mathbb{1} e^{ \frac{i}{\lambda} S[X,P,\Lambda]},\\
= \int^F_I [dX(t)dP(t)d\Lambda (t)]_{ _I}^{_F}\Delta_{FP}\int [d \epsilon] \delta[\mathcal{F}(\Lambda(\epsilon)]e^{ \frac{i}{\lambda} S[X,P,\Lambda]},\\
=\int [d \epsilon] \int^F_I [dX(t)]_{ _I}^{_F}e^{ \frac{i}{\lambda} S[X,\dot{X},\Lambda=1]}\left(\det \left[ \frac{d}{dt} \right]\right) . \label{eq:gfpi}
\end{eqnarray}
We have isolated $Vol(gauge)$ which can now be dropped. The crucial Faddeev-Popov determinant factor will now cancel one unphysical degree of freedom in the path integral because there are still three coordinates $X^a$ but only two physical degrees of freedom. A slight tidying up of the previous results gives as a final form in either phase space or configuration space as
\begin{eqnarray}
\langle X_{F},t_{F} \vert X_{I},t_{I} \rangle &=& \int \frac{[dXdPd\Lambda d\epsilon]_{ _I}^{_F}}{\mathcal{G}}\Delta_{FP}\delta[\mathcal{F}(\Lambda(\epsilon)]  e^{ \frac{i}{\lambda} S[X,P,\Lambda]}, \nonumber \\ \label{eq:gfpi2}
\\
&=& \int \frac{[dXd\Lambda d\epsilon]_{ _I}^{_F}}{\mathcal{G}}\Delta_{FP}\delta[\mathcal{F}(\Lambda(\epsilon)] e^{ \frac{i}{\lambda} S[X,\dot{X},\Lambda]}. \nonumber \\ \label{eq:gfpi3}
\end{eqnarray}
The advantage of this approach is that for a more general refractive index, we can perform propagation along a general curve rather than just a straight line.

\section{Worldline representations and propagating degrees of freedom}
\label{sec:worldline}

The path integral picture presented thus far gives a prescription for propagation between two points as shown in Figure~\ref{fig:setup}. However, due care should be given to the two types of boundary restrictions we are confining the problem to:
\begin{itemize}
\item The case where we are considering propagation between two planes (Figure~\ref{fig:setup}) relevant to optical imaging, with two spatial degrees of freedom.
\item The case where the initial and final points are unconstrained, relevant to general radiation, with three spatial degrees of freedom.
\end{itemize}
In the former situation (of most interest to the current paper) Equation~(\ref{eq:gfpi3}) can be used directly to evaluate propagation. In essence, the Faddeev-Popov determinant will cancel the un-physical locally longitudinal mode associated with a ray solution of the classical equations of motion. This in turn allows a trade of the evolution parameter $t$ for a coordinate that parametrises the classical solution. It follows from this that the two physical degrees of freedom are propagated between the initial and final imaging planes.

In the latter case, with unconstrained initial and final points, it is useful to discuss how the familiar propagation kernel (Green's function) associated with the Helmholtz equation may be recast in a path-integral worldline form~\cite{kleinert2009path}. 

Consider the Helmholtz equation when the refractive index is constant. The corresponding propagation kernel $K(X,X^{\prime})$ is defined by
\begin{equation}
\left(\lambda^2\delta^{ab}\partial_{a}\partial_{b} +n^2\right)K(X,X^{\prime})=\delta^3(X-X^{\prime}). \label{eq:helmholtz2}
\end{equation}
This propagator can be written in a number of different representations. Firstly it may be written via Fourier transformation as
\begin{eqnarray}
K(X,X^{\prime}) =\int \frac{d^3P}{(2\pi)^3}\frac{e^{i P_a(X-X^{\prime})^a}}{P^2 - n^2}.
\end{eqnarray}
Secondly we may invert the Helmholtz operator such that
\begin{eqnarray}
K(X,X^{\prime}) = \langle X\vert \frac{1}{\left(\lambda^2\delta^{ab}\partial_{a}\partial_{b} +n^2\right)} \vert X^{\prime} \rangle . \\
\end{eqnarray}
The last form can also be formally rewritten as
\begin{eqnarray}
K(X,X^{\prime})&=& \int_0^{\infty} dT \langle X\vert e^{-T(\lambda^2\delta^{ab}\partial_{a}\partial_{b} +n^2)} \vert X^{\prime} \rangle ,\\
&=&\int_0^{\infty} dT \int^{X(T)}_{X(0)} [dX(t)]e^{\frac{i}{(2\lambda)}\int^{T}_{0} dt (\delta_{ab}\dot{X}^a\dot{X}^b+n^2)}. \nonumber \\
\label{eq:worldline}
\end{eqnarray} 
This last expression can be obtained from Equation~(\ref{eq:gfpi3}) in the following way. Firstly the gauge is fixed to $\Lambda (t) = 1$. Secondly, both the left hand side and the right hand side are integrated as follows:
\begin{eqnarray} 
T:=t_F-t_I, \\
\int_0^{\infty} dT \langle X_{F},t_{F} \vert X_{I},t_{I} \rangle  =  \langle X_{F} \vert X_{I} \rangle\\ 
= \int_0^{\infty} dT\int [dX]_{ _I}^{_F}\Delta_{FP} e^{ \frac{i}{\lambda} S[X,\dot{X},\Lambda = 1]}. \nonumber \\ \label{eq:gfpi4}
\end{eqnarray}	
Thus the propagator for the standard Green's function may be written in a worldline form~\cite{chaichian2001}~\footnote{The Faddeev-Popov determinant can be absorbed in the normalization after some technical considerations~\cite{chaichian2001}.}.

\section{Vector Path Integrals}

Turning attention now to vectorial electromagnetic fields, the key to understanding polarisation aspects is a proper implementation of its spatial vector indices. Previously, the scalar field $\Phi(X,t)$ and its propagator $\langle \Phi(X_{F},t_F) \vert \Phi(X_{I},t_F)\rangle$  has been written as a worldline path integral amplitude 
\begin{equation}
\langle \Phi(X_{F},t_F) \vert \Phi(X_{I},t_F) \rangle \rightarrow \langle X_{F},t_{F} \vert X_{I},t_{I} \rangle. \nonumber
\end{equation}
By extension, a vector field $\mathbf{\mathcal{V}}_{a} $ (where we include an evolution parameter $t$ as before) will correspondingly need a worldline path integral representation to be written as 
\begin{equation}
\langle \mathcal{V}_{a} (X_{F},t_F) \vert \mathcal{V}_{b} (X_{I},t_I) \rangle \rightarrow \langle X_{F},t_{F} \vert X_{I},t_{I} \rangle_{ab}.
\end{equation}
We see that the propagation kernel must now become a second rank tensor (matrix) valued object~\footnote{This is the same index structure as the Green's dyadic function that one commonly encounters in scattering problems.}. As before, using an auxillary einbein function will allow for evolution along the rays.

Starting from Maxwell's equations, we can decouple the electric $\mathbf{E}^a$ and magnetic $\mathbf{B}^a$ such that
\begin{eqnarray}
	(\lambda^2\partial \wedge \partial\wedge)_{ab} +n^2\delta_{ab})\mathbf{E}^{b} = 0, \\
	(\lambda^2\partial \wedge n^{-2}\partial\wedge)_{ab} +\delta_{ab})\mathbf{B}^b = 0.
\end{eqnarray}
The above vector field equations automatically satisfy the remaining Maxwell's equations, $\partial \cdot \mathbf{B}=0$, and $\partial \cdot (n^2\mathbf{E})=0 $, as well. It is worth remarking here that we are making a restriction on the constitutive relations of the electromagnetic fields to not include chiral matter, nor birefringence, that may be spatially varying. To simplify the presentation and general scheme of the approach, these inclusions are left for future work. Thus the two vector differential operators are necessarily transverse ab initio. The wedge products can be simplified to give 



\begin{eqnarray}
	\left\{ (\lambda^2\partial^2 +n^2)\delta_{ab}+(\lambda\partial_{b} \ln n^2)(\lambda\partial_{a})\right. \nonumber \\
\left.+ (\lambda^2\partial_{a}\partial_{b} \ln n^2)\right\}\mathbf{E}^{b} = 0,\label{eq:electricfe} \\ 
\triangle^{E}_{ab}\mathbf{E}^{b} \equiv 0, \\
\left\{(\lambda^2\partial^2 +n^2)\delta_{ab}+(\lambda\partial_{b} \ln n^2)(\lambda\partial_{a})\right. \nonumber \\
\left. -\delta_{ab}(\lambda\partial_{c} \ln n^2)(\lambda\partial_{c} )\right\}\mathbf{B}^{b} = 0, \label{eq:magneticfe} \\
\triangle^{M}_{ab}\mathbf{B}^{b} \equiv 0. 
\end{eqnarray}

These decoupled vector equations of motion are interesting from the differential geometry standpoint~\cite{misner1973gravitation}, in terms of a metric $g_{ab}$, and in its relation to the scalar Helmholtz equation. Consider the following points. Firstly, there are derivatives of the refractive index here that are not present in the scalar Helmholtz equation. A better scalar model could include these terms by for example promoting partial to covariant derivatives for the scalar field, thereby obtaining linear connection terms~\cite{misner1973gravitation}~\footnote{The connection here would be given in terms of the metric $g_{ab}=n^2\delta_{ab}$.}. Secondly, electric and magnetic fields behave differently in a dielectric medium and so a naive choice of metric $g_{ab}=n^2\delta_{ab}$ isn't sufficient to describe them both as vector fields at the same time. Thirdly, minimal substitutions where we replace a partial derivative with a covariant derivative produces extra quadratic (nonlinear) terms in the connection that are not present in Maxwell's equations.

The crucial observation to make here for the connection with the path integral is that a scalar Hamiltonian will be insufficient to capture vectorial aspects. The corresponding optical path length consequently must be generalised to a \emph{matrix}~. Equations~(\ref{eq:electricfe}) and~(\ref{eq:magneticfe}) allow us to define the electric and magnetic tensor Hamiltonian operators as
\begin{eqnarray}
	\hat{H}^{E}_{ab} \equiv \lambda \Lambda (t) \triangle^{E}_{ab}, \\
	 \hat{H}^{M}_{ab} \equiv \lambda \Lambda (t) \triangle^{M}_{ab}.
\end{eqnarray}
As with the scalar field, we now consider the lifted fields $\mathcal{E}^a(X,t)$ and $\mathcal{B}^a(X,t)$ that satisfy matrix Schr\"{o}dinger equations
\begin{eqnarray}
\hat{H}^{E}_{ab}(X,\partial, \Lambda)\mathcal{E}^b(X,t)= i\lambda\frac{\partial \mathcal{E}_a(X,t)}{\partial t}, \\
\hat{H}^{M}_{ab}(X,\partial, \Lambda)\mathcal{B}^b(X,t)= i\lambda\frac{\partial \mathcal{B}_a(X,t)}{\partial t}.
\end{eqnarray} 
The corresponding OP functional for vector fields is derived in direct analogy to the action given in Equation~(\ref{eq:action}) for scalar fields. The key difference here is that the optical path length must be generalised to a matrix (in general tensor) valued functional:-
\begin{eqnarray}
	S[X,\dot{X},\Lambda] &\rightarrow & S_{ab}[X,\dot{X},\Lambda].
\end{eqnarray}
A non-trivial aspect now enters in terms of the conjugate momentum. The standard quantum prescription of identifying $P_a \equiv i\lambda\partial_a$ as in~\cite{Gloge:69}, where the differential operator is identified with a $c$-number has a different relation to velocity. This is because we need to transport both momentum \emph{and} spin (or polarization) along the curve $X^a(t)$. In addition, because we have two different Hamiltonians for the $\mathbf{E}^a$ and $\mathbf{B}^a$ fields, we require two distinct OP functionals for each field
\begin{eqnarray}	
S^E_{ab}[X,P,\Lambda]& = & \int^{t_F}_{t_I} dt P_a\dot{X}_b-H^{E}_{ab}, \label{eq:electricOP}  \\
S^M_{ab}[X,P,\Lambda]& = & \int^{t_F}_{t_I} dt P_a\dot{X}_b-H^{M}_{ab},
\end{eqnarray}
where
\begin{eqnarray}
H^{E}_{ab} &=& \frac{\lambda \Lambda (t)}{2}[(P^2  -n^2)\delta_{ab} +2i\lambda \Gamma_{b}P_{a}+ \lambda^2\partial_{a}\Gamma_{b}], \label{eq:electricH} \\ 
H^{M}_{ab} &=& \frac{\lambda \Lambda (t)}{2}[(P^2  -n^2)\delta_{ab}+2i\lambda \Gamma_{b}P_{a} -2i\lambda \Gamma_{c}P_{c}\delta_{ab}],\label{eq:magneticH} \\
\Gamma_a &:=& \frac{1}{2}\partial_{a} \ln n^2 . 
\end{eqnarray}
A similar procedure was used in the case of spin-$1/2$ particles in~\cite{kleinert2009path}. Note also the similarity to the path integral formulated on curved spaces where the derivative is modified to include a connection term~\cite{schulman2005techniques}.

With these matrix valued OP functionals substituted into the Feynman path integral, one can obtain the vector equivalent of the scalar field propagator. It is given by
\begin{eqnarray}
\langle X_{F},t_{F} \vert X_{I},t_{I} \rangle^{E}_{ab} = \int \frac{[dXdPd\Lambda d\epsilon]_{ _I}^{_F}}{\mathcal{G}}\Delta_{FP}\delta[\mathcal{F}(X^a(\epsilon)]  \nonumber \\
\times \mathcal{P}\left\lbrace e^{ \frac{i}{\lambda} S^{E}_{ab}[X,P,\Lambda]}\right\rbrace, \\
  \label{eq:pielectric}
\langle X_{F},t_{F} \vert X_{I},t_{I} \rangle^{M}_{ab} = \int \frac{[dXdPd\Lambda d\epsilon]_{ _I}^{_F}}{\mathcal{G}}\Delta_{FP}\delta[\mathcal{F}(X^a(\epsilon)]  \nonumber \\
\times \mathcal{P}\left\lbrace e^{ \frac{i}{\lambda} S^{M}_{ab}[X,P,\Lambda]}\right\rbrace,
  \label{eq:pimagnetic}    
\end{eqnarray}
The symbol $\mathcal{P}$ denotes path ordering because the exponential is matrix valued and therefore in general non-commutative, whilst the superscript $E(M)$ indicates whether it is the electric or magnetic field we are evaluating~\footnote{We will will define more precisely the path ordered exponential in Section \ref{sec:eikonaldg}}. Note also some similarities with~\cite{Gersten:87}, but in this work the crucial Faddeev-Popov factors are not present that ensure the right integration measure.

Moving to configuration space by integrating over momenta, one finds
\begin{eqnarray}
\langle X_{F},t_{F} \vert X_{I},t_{I} \rangle^{E}_{ab} = \int \frac{[dXd\Lambda d\epsilon]_{ _I}^{_F}}{\mathcal{G}}\Delta_{FP}\delta[\mathcal{F}(X^a(\epsilon)]  \nonumber \\
\times \mathcal{P}\left\lbrace e^{ \frac{i}{\lambda} S^{E}_{ab}[X,\dot{X},\Lambda]}\right\rbrace,
  \label{eq:pielectric2} \\
\langle X_{F},t_{F} \vert X_{I},t_{I} \rangle^{M}_{ab} = \int \frac{[dXd\Lambda d\epsilon]_{ _I}^{_F}}{\mathcal{G}}\Delta_{FP}\delta[\mathcal{F}(X^a(\epsilon)]  \nonumber \\
\times \mathcal{P}\left\lbrace e^{ \frac{i}{\lambda} S^{M}_{ab}[X,\dot{X},\Lambda]}\right\rbrace,
  \label{eq:pimagnetic2}    
\end{eqnarray}
where
\begin{eqnarray}	
	S^E_{ab}[X,\dot{X},\Lambda]& = & \frac{1}{2}\int^{t_F}_{t_I} dt \left(\frac{1}{\lambda\Lambda (t)}\dot{X}_c \dot{X}^c+\lambda \Lambda (t)n^{2}\right)\delta_{ab} \nonumber \\
	&&+i\lambda \Gamma_{b}\dot{X}_{a}+\lambda^2\partial_{a}\Gamma_{b}, \label{eq:electricOP2} \\
	&=& S[X,\dot{X},\Lambda]\delta_{ab}+i\lambda\int^{t_F}_{t_I} dt \Gamma_{b}\dot{X}_{a} \nonumber \\
	&&+\mathcal{O}(\lambda^2), \label{eq:electricOP3}\\
	S^M_{ab}[X,\dot{X},\Lambda]& = & \frac{1}{2}\int^{t_F}_{t_I} dt \left(\frac{1}{\lambda\Lambda (t)}\dot{X}_c \dot{X}^c+\lambda \Lambda (t)n^{2}\right)\delta_{ab} \nonumber \\
	&&+i\lambda \Gamma_{b}\dot{X}_{a}-i\lambda \Gamma_{c}\dot{X}_{c}\delta_{ab}, \label{eq:magneticOP2} \\
	&=& S[X,\dot{X},\Lambda]\delta_{ab}+i\lambda\int^{t_F}_{t_I} dt (\Gamma_{b}\dot{X}_{a}-\Gamma_{c}\dot{X}^{c}\delta_{ab}) \nonumber \\
	&&+\mathcal{O}(\lambda^2).\label{eq:magneticOP3}
\end{eqnarray}
The last term in Equation~(\ref{eq:magneticOP2}) is a total derivative of the $t$ parameter and as such only depends on the initial and final conditions. So both the electric and magnetic OP functionals are the same up to higher derivative corrections (and therefore higher order in $\lambda$) for the electric field.

\section{Differential geometry, the matrix optical path length and the eikonal}
\label{sec:eikonaldg}

It is interesting to consider how this matrix picture for the optical path length arises in the short wavelength limit. Standard expressions for the electric and magnetic fields written in terms of the eikonal function and the amplitude vectors can be found in~\cite{bornwolf2020}. As an expansion in wavelength to quadratic order, one finds (i) the eikonal equation for the eikonal at zeroth order, (ii) the first order transport equations for the amplitude vectors at linear order, and (iii) second order transport equations for the amplitude vectors at quadratic order. When the eikonal satisfies $\partial^a \mathcal{S}=\dot{X}^a(t)$ then two differential geometric facts follow. Firstly, rays follow geodesic paths $X^a(t)$. Secondly, the amplitude vectors (more properly when they have been normalised to unity) are parallel transported along these geodesic curves $X^a(t)$~\cite{bornwolf2020}. 

Standard expressions for the electric and magnetic fields written in terms of the eikonal function $S$ and the two amplitude vectors $\mathbf{e}$ and $\mathbf{b}$ are  
\begin{eqnarray}
\mathbf{E}^a(X) & = & \mathbf{e}^a(X)e^{\frac{2\pi i}{\lambda}\mathcal{S}(X)}, \label{eq:eikonalE1} \\
\mathbf{B}^a(X) & = & \mathbf{b}^a(X)e^{\frac{2\pi i}{\lambda}\mathcal{S}(X)}, \label{eq:eikonalM1} \\
u^a& \equiv & \mathbf{e}^a/(\vert \mathbf{e} \vert), \\
v^a& \equiv & \mathbf{b}^a/(\vert \mathbf{b} \vert).
\end{eqnarray}
A key set of standard results described in~\cite{bornwolf2020} for eikonal solutions of the electric $\mathbf{E}(X)$ and magnetic  $\mathbf{B}(X)$ fields are the following. There are three geometric facts that (i) rays follow geodesic paths $X^a(t)$, (ii) the polarization vectors are parallel transported along these geodesic curves $X^a(t)$ and (iii) they are mutually orthogonal. It is also stated there~\cite{bornwolf2020} that the differential geometric version of this is given by the parallel transport of the polarization vector with respect to an affine connection. The connection used is the Levi-Civita connection $\Gamma^{a}_{bc}$ (commonly used in general relativity~\cite{misner1973gravitation}) which can be expressed in terms of a metric $g_{ab}$. In the current system, these can be written as
\begin{eqnarray}
	g_{ab} &=& n^2\delta_{ab}, \label{eq:metric} \\
	\Gamma^{a}_{bc} &=& \frac{1}{2}g^{ad}(\partial_c g_{db} + \partial_bg_{dc} - \partial_d g_{bc}), \\
	&=& \frac{1}{2}(\delta^{a}_{b}\partial_c \ln n^2 + \delta^{a}_{c}\partial_b \ln n^2 -\delta_{bc}\partial^a \ln n^2).
\end{eqnarray}
The connection allows one to perform covariant differentiation on a curved space with the metric $g_{ab}$ and in turn to define parallel transport of vector fields. The covariant derivative $\nabla_a$ of a vector field $V^b$ is given by
\begin{eqnarray}
	\nabla_a V^b \equiv \partial_aV^b+\Gamma^{b}_{ac}V^c.
\end{eqnarray}
The parallel transport of a vector field $V^b$ along the rays curve $X^{a}(t)$ is then simply
\begin{eqnarray}
	D_{\dot{X}}V^a \equiv \dot{X}^c\nabla_c V^a = \partial_tV^a+\Gamma^{a}_{bc}V^b\dot{X}^c=0. \label{eq:parallelT}
\end{eqnarray}
Equation~(\ref{eq:parallelT}) yields the parallel transport of amplitude vectors (given in~\cite{bornwolf2020}) when $V^a$ is a polarization vector $u^a$ or $v^a$ and gives the geodesic equation for rays when $V^a=\dot{X}^a$. What the parallel transport equations show are the differential geometric side to ray and wave propagation. Indeed, Equation~(\ref{eq:parallelT}) may be recast as a path-ordered exponential~\cite{Carroll:1997ar} already encountered as
\begin{eqnarray}
V^a (t_{F}) = \mathcal{P}\left\{ \exp \left(-\int^{t_F}_{t_I}dt \Gamma^{a}_{bc}\dot{X}^c \right) \right\}V^b (t_{I}). \label{eq:parallelT2}
\end{eqnarray}
The path ordered exponential for a matrix exponent $M^a_b(t)$, is defined by its power series expansions as~\footnote{This is basically the time ordering operation of hamiltonians that occurs in the interaction picture in quantum mechanics but now as a matrix.}
\begin{eqnarray}
V^a (t_{F}) &=& \mathcal{P}\left\{ \exp \left(-\int^{t_F}_{t_I}dt M^{a}_{b}(t) \right) \right\}V^b (t_{I}) \\
&=& \sum_{m=0}^{\infty}\frac{(-1)^m}{m!}\int^{t_F}_{t_I}dt_m \cdots \int^{t_F}_{t_I}dt_1\mathcal{P}\left\{M^{a}_{c}(t_m)\cdots \right. \nonumber \\
&&\left. \cdots M^{d}_{b}(t_1)\right\}V^b (t_{I}) \\
 &:=&  \left\{ \delta^a_b+ \left(-\int^{t_F}_{t_I}dt_1 M^{a}_{b}(t_1) \right) \right. \nonumber \\ 
&&+ \left. \left(\int^{t_F}_{t_I}dt_2\int^{t_2}_{t_I}dt_1 M^{a}_{c}(t_2) M^{c}_{b}(t_1) \right) \right. \nonumber \\
&&\left. +\cdots \right\}V^b (t_{I}).  \label{eq:parallelT3}
\end{eqnarray}
The integral and differential forms of parallel transport (Equations~(\ref{eq:parallelT}) and (\ref{eq:parallelT2})) can be further elucidated by using the conformal metric, Equation~(\ref{eq:metric}). Assuming that the vector is orthogonal to the tangent vector, $V_c\dot{X}^c = 0$,  one finds
\begin{eqnarray}
\frac{d (n V^a)}{dt} &=& -\dot{X}^a (nV^b)\partial_{b} \ln n , \\
	n(t_F)V^a (t_{F}) &=& \mathcal{P}\left\{ \exp \left(-\int^{t_F}_{t_I}dt \dot{X}^a \partial_b \ln n \right) \right\}n(t_I)V^b (t_{I}) \nonumber \\
	&=& \mathcal{P}\left\{ \exp \left(-\int^{t_F}_{t_I}dt \dot{X}^a \Gamma_b  \right) \right\}n(t_I)V^b (t_{I}). 
\end{eqnarray}
This is exactly the form of parallel transport given in the vector path integrals Equations~(\ref{eq:pielectric2}) and~(\ref{eq:pimagnetic2}). Actually this can be simply related to the the group $SO(3)$ by writing out $\partial_a n^2(X)$ explicitly, viz
\begin{eqnarray}
	V_a (t_{F}) &=& \beta\mathcal{P}\{ \exp \left(-\int^{t_F}_{t_I}dt \alpha(X)(\dot{X}_a X_b \right. \nonumber \\
&& \left.  -\dot{X}_b X_a) \right) \}V^b (t_{I}),\\
&=& \beta\mathcal{P}\left\{ \exp \left(-\int^{t_F}_{t_I}dt \alpha(X)J_{ab}  \right) \right\}V^b (t_{I}),
\end{eqnarray}
where $J_{ab}$ are the Lie algebra generators of the group $SO(3)$, $\beta$ is a numerical constant and $\alpha (X)$ is local function of $X$~\footnote{The second term introduced here is normally zero because $V_c\dot{X}^c = 0$. When this is not the case the second term must be retained.}. Thus the parallel transport of the polarisation vectors is equivalent to a local rotation in three dimensions, where the derivative of the refractive index gives a local rotation angle. Because the Lie algebra elements are non-commutative, this gives a clear reason for the use of a path-ordered exponential. This may also be clearly seen before integrating over momenta in the path integral where in the above we would just substitute $P_a$ for $\dot{X}^a$ whence $J_{ab}$ just becomes the usual angular momentum operator.

How does the previous matrix model for vector path integrals fit in with parallel transport found in standard eikonal theory? Recall that normally vector fields $\mathbf{V}^a(X)$ are decomposed as a vector amplitude $V^a(X)$ and a \emph{scalar} phase function $\mathcal{S}(X)$ as $\mathbf{V}^a(X)=e^{\frac{ i}{\lambda}\mathcal{S}(X)}V^a(X)$. However, it is also possible to write the phase as a \emph{matrix} (tensor) valued function rather than a scalar, viz
\begin{eqnarray}
\mathbf{E}_{a}(X) = \mathcal{P}\left\lbrace e^{\frac{ i}{\lambda}\mathcal{S}_{ab}(X)}\mathbf{u}^b(X)\right\rbrace, \label{eq:eikonalE} \\
\mathbf{B}_{a}(X) = \mathcal{P}\left\lbrace e^{\frac{ i}{\lambda}\mathcal{S}_{ab}(X)}\mathbf{v}^b(X)\right\rbrace . \label{eq:eikonalB}
\end{eqnarray}
The origin of this can be seen by returning to Equations~(\ref{eq:pielectric2}) - (\ref{eq:magneticOP2}) where they can be evaluated semi-classically~\cite{Berry_1972} to be
\begin{eqnarray}
\langle X_{F},t_{F} \vert X_{I},t_{I} \rangle_{ab} = & \mathcal{P}\left\lbrace e^{-\int^{t_F}_{t_I} dt \dot{\mathcal{X}}_{a}\Gamma_{b} + \frac{i}{\lambda} S_{ab}[\mathcal{X},\dot{\mathcal{X}}]  }\det (\mathcal{D}) \right\rbrace ,
\end{eqnarray}
where $\det (\mathcal{D})$ is essentially the Van-Vleck determinant and $X^a(t)=\mathcal{X}^a(t)$ is a solution of the classical equations of motion ($\Lambda$ has been fixed as per the discussion of the last section). We see that both the eikonal and parallel transport aspects are captured with this matrix approach.

An interesting comparison may be made with~\cite{Dimant:10} where using a spin coherent states approach for the polarisation, the Serret-Frenet equations are found for the polarisation vectors in the short wave limit.

\section{Applications by example}

In the following subsections, explicit examples are given for both the scalar field case and the vector field. These use the path integral in a practical sense and serve to illustrate how by using a ray formulation of optics as a path integral, many wave optics phenomena can be recovered or uncovered.

\subsection{Example 1: Gouy phase for the scalar field}
\label{sec:example1}

\begin{figure}
\begin{center}
\includegraphics[scale=0.6]{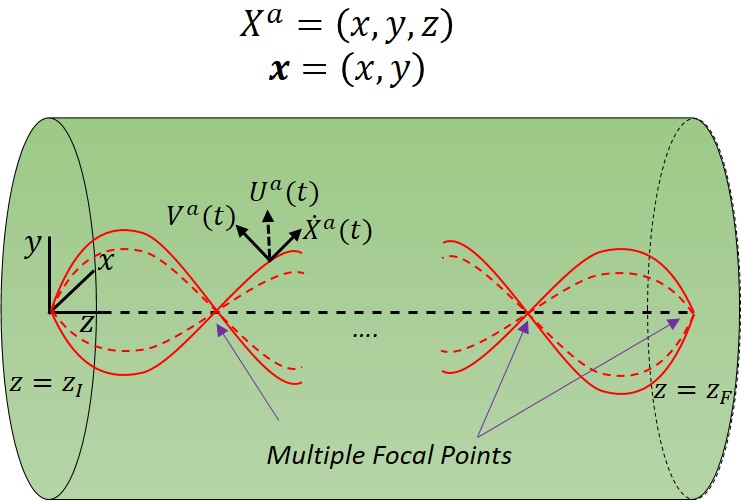} 
\caption{A quadratic GRIN fiber rod and the discrete multiple lens relay.} \label{fig:relay}
\end{center} 
\end{figure}

In this first example application, we will consider how the Gouy phase manifests itself for light propagation in a media that admits multiple focal points. The system we consider is a selfoc (self focusing) quadratic GRIN media that can be used to model fibers, with the relevant details shown in Figure~\ref{fig:relay}. This media acts as a relay imaging optic because the ray solutions~\cite{Marchand:72} admit multiple intermediate image planes. It is a natural extension of our previous results~\cite{Babington:18}, since there the focal points were not explored. Of note here is also the work~\cite{Alieva:08}, where the Gouy phase was considered in the context of the fractional Fourier transform. 

The refractive index profile is given by
\begin{eqnarray}
n^2(\mathbf{x},z)=n_0^2(1-\mathbf{x}^2/R^2), \label{eq:grin}
\end{eqnarray}
where $X^a=(\mathbf{x},z)$. This profile admits exact ray solutions~\cite{Marchand:72} of the equations of motion derived from Equation~(\ref{eq:opl}) in the form of sinusoids, together with perfect focussing of the rays. Thus the classical solutions are well determined and will be used subsequently in evaluating the path integral.

Turning now to the corresponding wave aspects for the media given by Equation~(\ref{eq:grin}), from Figure~\ref{fig:relay} the imaging problem is clear. We are propagating two spatial degrees of freedom between an initial and final plane defined by $z=z_{I}$ and $z=z_{F}$. One can evaluate the path integral Equation~(\ref{eq:gfpi3}) by the following prescription derived from Section~\ref{sec:review}:
\begin{enumerate}
\item Fix the reparametrization invariance by choosing a gauge $\Lambda = 1$.
\item Remove the spurious degree of freedom in the path integral by use of the Faddeev-Popov determinant. 
\end{enumerate}  
Since $n^2$ is independent of the $z$-coordinate, we can evaluate the free space $z$-coordinate path integral directly as a determinant~\cite{chaichian2001}. By substituting gauge choice into Equation~(\ref{eq:gfpi3}) and using the form of the Faddeev-Popov determinant Equation~(\ref{eq:fpdet2}) we find
\begin{eqnarray}
\langle X_{F},t_{F} \vert X_{I},t_{I} \rangle 
= \int^F_I [dX(t)]_{ _I}^{_F}e^{ \frac{i}{\lambda} S[X,\dot{X},\Lambda=1]}\left(\det \left[ \frac{d}{dt} \right]\right), \label{eq:piho} \\
S[X,\dot{X}, \Lambda =1] = \frac{1}{2\lambda}\int^{t_F}_{t_I} dt \left(\dot{\mathbf{x}}^2+\dot{z}^2 
 +\lambda^2 n_0^2(1-\mathbf{x}^2/R^2)\right).\label{eq:pihoa}
\end{eqnarray}
The path integral then factorises into a separate one dimensional longitudinal part and two dimensional transverse part. The $z$-path integral can be performed by making the substitution
\begin{eqnarray}
z(t)=z(t)_{class.}+Z(t).
\end{eqnarray}
This gives rise to two features. Firstly, the classical solution $z(t)_{class.}$ can be used to eliminate $t$ in favour of $z$. The classical $z$ equations of motion (together with an initial condition and constraint equation) becomes
\begin{eqnarray}
z(t)_{class.} = (t-t_I )\lambda n_0 +z_I. \label{eq:zclass}
\end{eqnarray}
Secondly, the remaining $Z(t)$-integral can be evaluated as a determinant, viz
\begin{eqnarray}
\int [dZ(t)] \exp \left(\frac{i}{2\lambda^2} \int dt \dot{Z}^2 \right) = \left(\det\left[ -\frac{d}{dt}^2\right]\right)^{-1/2}.
\end{eqnarray}
This is precisely cancelled by the Faddeev-Popov factor in Equation~(\ref{eq:piho})~\footnote{By using the operator identity $\det A^2=\det A\det A$}. Now the path integral takes the reduced form
\begin{eqnarray}
\langle \mathbf{x}_{F},z_{F} \vert \mathbf{x}_{I},z_{I} \rangle 
= \int [d \mathbf{x}(t)]_{ _I}^{_F} e^{ \frac{i}{2\lambda}S[\mathbf{x},\dot{\mathbf{x}}]}, \label{eq:piho2}
\end{eqnarray}
where the reduced action is
\begin{eqnarray}
S[\mathbf{x},\dot{\mathbf{x}}]=\frac{n_0}{2}\int^{z_F}_{z_I} dz\left(\dot{\mathbf{x}}^2 -\mathbf{x}^2/R^2 \right). \label{eq:hoaction}
\end{eqnarray}
In the above we have relabelled $\dot{\mathbf{x}}=d\mathbf{x}/dz$ and used Equation~(\ref{eq:zclass}) to rescale the $t$-parameter. This is just the action for a two dimensional simple harmonic oscillator. The path integral has a well known analytic form which is
\begin{eqnarray}
\langle \mathbf{x}_{F},z_{F} \vert \mathbf{x}_{I},z_{I} \rangle = \frac{1}{(2\pi i)}\left( \frac{n_0}{\lambda R\sin (z_F-z_I)/R} \right) \nonumber \\
\times \exp \left[ \frac{in_0}{4\lambda R \sin (z_F-z_I)/R}\right. \nonumber \\
\left. \times [(\mathbf{x}^2_{F}+\mathbf{x}^2_{I}) \cos (z_F-z_I)/R-2\mathbf{x}_{F}\mathbf{x}_{I} ]\right]. \label{eq:amplitude}
\end{eqnarray}
This is an interesting formula not just because it is an analytic solution. It is the equivalent of the Fresnel diffraction propagator in the case of a focussing media, where the focal points are given by
\begin{eqnarray}
(z_F-z_I)/R = m\pi, \label{eq:focal}
\end{eqnarray}
with $m \in \mathbf{Z}$. This gives the number of focal points the field passes through. Also note that we haven't taken any far field limit or made a paraxial approximation. The reader should consult~\cite{Gloge:69} in particular as in that paper the same GRIN media is considered in the paraxial limit. 

It is clear from Equation~(\ref{eq:focal}) that something interesting happens at the focal points where the amplitudes blow up. That the propagator diverges isn't neccesarilly a problem as this is a distribution and will correspond to a well behaved field. However, it is less clear what happens to the phase of the propagator whilst passing through a focal point. In fact Equation~(\ref{eq:amplitude}) is valid provided we do not pass through a focal point. A more careful treatment in evaluating the path integral leads to the following Feynman-Soriau form~\cite{Souriau1976}
\begin{eqnarray}
\langle \mathbf{x}_{F},z_{F} \vert \mathbf{x}_{I},z_{I} \rangle = Det (S) \exp [ i\Delta_f+i\Delta_{n}], \\
Det (S)= \left( \frac{n_0}{2\pi\lambda R\mid\sin (z_F-z_I)/R\mid} \right) \\
\Delta_n = \left(\frac{n_0}{4\lambda R  \sin (z_F-z_I)/R}\right) \nonumber \\
\times \left[(\mathbf{x}^2_{F}+\mathbf{x}^2_{I}) \cos (z_F-z_I)/R-2\mathbf{x}_{F}\mathbf{x}_{I} \right], \label{eq:hoprop} \\
\Delta_f = -\frac{\pi}{2}\left(2+2 m \right).
\label{eq:amplitude2}
\end{eqnarray}
The expression for $\Delta_f$ involves two pieces. The first phase factor results from a careful treatment of $\sqrt{i}$. The second piece is the Maslov-Morse~\cite{chaichian2001} index that counts multiplicities when rays pass through a focal point. The integer $m$ is derived via the floor function $\mathbf{Floor} [(z_F-z_I)/R] = m$, where the floor function rounds down to the lowest integer e.g. $\mathbf{Floor} [3/2] = 1$ etc. In our example the integer $m$ counts how many focal points in the GRIN relay have been passed through, and the multiplier of 2 corresponds to a dimension two focal point. These are exactly the two degrees of freedom the path integral encodes as well as it being degenerate due to the harmonic potential being axially symmetric. Thus passing through a single focal point one accrues a phase of $\pi$ which is sufficient to observe destructive interference.

\begin{figure}
\begin{center}
\includegraphics[scale=0.4]{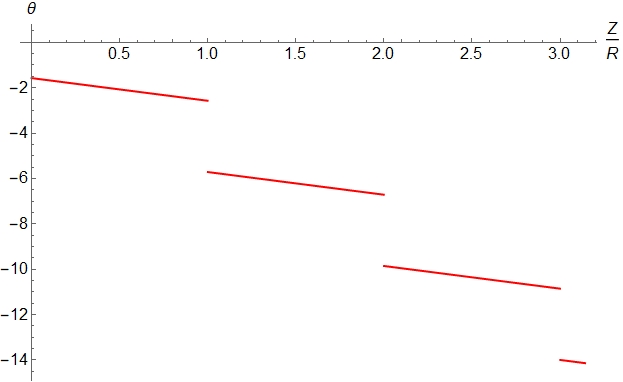} 
\caption{Gouy phase for an initial Gaussian beam wavepacket that is a ground state eigen-function of the reduced form path integral.} \label{fig:gouyphase}
\end{center} 
\end{figure}

This relates directly to the Gouy phase. We have an explicit expression for the propagator that can now be used to propagate an initial field configuration along the $z$-direction. Consider a standard Gaussian wavepacket as the initial field given by
\begin{eqnarray}
\psi(\mathbf{x}_{I},z_{I})=\frac{1}{(2\pi \mathcal{A})^{1/2}}\exp \left( -\mathbf{x}_{I}^2/4\mathcal{A} \right).
\end{eqnarray}
It is well known that the Gaussian solution remains Gaussian as it evolves. Using the explicit form of the propagator in Equation~(\ref{eq:hoprop}) the propagated field is found to be
\begin{eqnarray}
\psi(\mathbf{x}_{F},z_{F})&=&\frac{1}{(2\pi \mathcal{F})^{1/2}}\exp \left( -\mathbf{x}_{F}^2/4\mathcal{F} \right)e^{i\theta},  \\
\mathcal{F} &:=& \mathcal{A}\left[\cos^2 \left(\frac{z_{F}-z_{I}}{R}\right)  + \frac{2\mathcal{A}n_0}{\lambda R}\sin^2 \left(\frac{z_{F}-z_{I}}{R}\right)\right], \nonumber \\
\\
\theta &=& \tan ^{-1}\left( \frac{2\mathcal{A}n_0}{\lambda R}\cot (z_{F}-z_{I})/R) \right) \nonumber \\
&& -\frac{n_0  \mathbf{x}_{F}^2}{2 \lambda R }\cot (z_{F}-z_{I})/R)\left(\frac{\mathcal{A}}{\mathcal{F}}-1\right) \nonumber \\
&& -\frac{\pi}{2}\left(2+2m \right),
\end{eqnarray} 
where $\theta$ is precisely the Gouy phase. In the case where the initial wavepacket is the zero eigen-function with $2\mathcal{A}n_0/\lambda R = 1$, then $\mathcal{F}=\mathcal{A}$ and the Gouy phase becomes particularly simple (taking $z_{I}=0$ and $z_{F}=z$ )
\begin{eqnarray}
\theta &=&   \frac{z}{R} -\pi\left(1+m \right).
\end{eqnarray}
It accounts for the periodic behaviour of the phase as the number of focal points traversed increases as shown in Figure~\ref{fig:gouyphase}. The infinite domain ($-\infty < z < +\infty$) of the usual Gouy phase profile is made finite ($-\pi R/2 < z < +\pi R/2$) by the periodic geodesics. Thus collimated light becomes focused and then re-collimated over a length of $\pi R$.

\subsection{Example 2: Vector fields in quadratic GRIN media}
\label{sec:example2}

As an extension application, we consider the previous GRIN media in the full vector theory (see also~\cite{2019NatCo} for practical applications). This is is instructive to see how the evaluation of the matrix path integral can proceed.

\begin{figure}
\begin{center}
\includegraphics[scale=0.4]{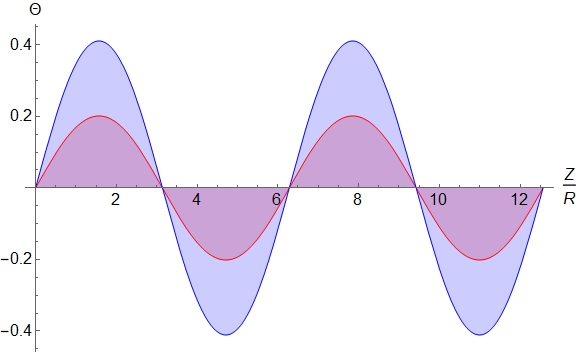} 
\caption{The longitudinal component of field propagation as described by parallel propagation along sinusoidal geodesics.} \label{fig:longitudinal}
\end{center} 
\end{figure}

Starting with the OP for the electric field given by Equation~(\ref{eq:electricOP2}), we fix the reparametrization invariance as in Section~\ref{sec:example1} with $\Lambda = 1$. To simplify the presentation, we consider a two dimensional source where the electric field is radially polarized. Writing out the non-zero components explicitly in cylindrical polar coordinates $(r,\phi, z)$ and using Equation~(\ref{eq:pihoa}) one finds
\begin{eqnarray}
S^E_{rr}[X,\dot{X}] &=& S[X,\dot{X},\Lambda = 1] +i\lambda\int^{t_F}_{t_I} dt  \Gamma_{r}\dot{r}, \\
S^E_{zr}[X,\dot{X}] &=& S[X,\dot{X},\Lambda = 1] +i\lambda\int^{t_F}_{t_I} dt  \Gamma_{r}\dot{z},
\end{eqnarray}
and similarly for the magnetic version. These admit the same harmonic oscillator solutions as in Section~\ref{sec:example2}, but in addition we have additional contributions due to the parallel propagation $\Gamma_a$ terms. Both of these can be evaluated explicitly along the geodesics given by the simple harmonic oscillator solutions
\begin{eqnarray}
\int^{t_F}_{t_I} dt  \Gamma_{r}\dot{r} &=& \int^{t_F}_{t_I} dt  \dot{r} \partial_r\ln n =\ln \frac{n(t_F)}{n(t_I)}, \label{eq:rfieldpt} \\
\int^{t_F}_{t_I} dt  \Gamma_{r}\dot{z} &=& \int^{t_F}_{t_I} dt  \dot{z} \partial_r\ln n =\int^{z_F}_{z_I} dz   \frac{-2r/R^2}{1-r^2/R^2}.\label{eq:zfieldpt}
\end{eqnarray}
The transverse component Equation~(\ref{eq:rfieldpt}) just depends on the end points, whereas the longitudinal component Equation~(\ref{eq:zfieldpt}) depends on the geodesic path followed. To be specific, consider the family of geodesics given by:
\begin{eqnarray}
r(z) &=& r_I\cos z/R,\\
z_I &=& 0,\\
\kappa &:=& r_I/R. 
\end{eqnarray}
Then Equation~(\ref{eq:zfieldpt}) can be explicitly evaluated to be:
\begin{eqnarray}
\int^{t_F}_{t_I} dt  \Gamma_{r}\dot{z} = \frac{-2}{(1-\kappa^2)^{1/2}}\tan^{-1}\left(\frac{\sin z/R}{(1-\kappa^2)^{1/2}}\right) :=\Theta .
\end{eqnarray}
This behaviour is shown in Figure~\ref{fig:longitudinal}. Thus the longitudinal variations display a similar periodic behaviour to the Gouy phase as they should do.

\subsection{Example 3: Basic Phenomenology}

To get a feel for some numbers, we use some recently published GRIN material data given in~\cite{ocier2020} that operates in the visible regime. We use the following parameters: 
\begin{itemize}
\item A design wavelength $\lambda = 633 nm$.
\item Refractive index variation of $n_{centre} = 1.8$ and $n_{edge}=1.6$ across a $20\mu m$ diameter fibre face.
\end{itemize}
These are used to derive the following optical parameters given in Table \ref{tab:grinparams}.
\begin{table}[htbp]
\centering
\caption{\bf Derived optical parameters for GRIN material}
\begin{tabular}{|c|c|}
\hline
Optical parameter & Value   \\
\hline
$R$ & $22\; \mu m$   \\
\hline
$n_{0}$ & $1.8$  \\
\hline
$\mathcal{A}$ & $3.84\; \mu m^2$   \\
\hline
\end{tabular}
  \label{tab:grinparams}
\end{table}
The values are useful in establishing possible applications and phenomenology. For example, one can see it is possible to construct optical interference devices on the tens of microns length scale. By using two GRIN fibres of differing length, one can arrange for each fibre to contain an odd or even number of focal points thereby producing destructive interference upon recombination. Another use could be basic phenomenology. The ratio $\lambda/R \approx 0.03$ is a meaningful perturbation parameter. This can be used to better understand the higher order terms neglected in Equations~(\ref{eq:electricOP3}) and~(\ref{eq:magneticOP3}) via standard perturbation theory.

\section{Conclusions}
\label{sec:conclusion}

To summarise, in this paper we have constructed a path integral representation of electromagnetic vector fields in a GRIN background refractive index. A key aspect in this approach is in using an action that is matrix rather than scalar valued. This leads to the use of an optical path length matrix, rather than the more familiar scalar function. The phase space path integrals that are initially gauge invariant can be gauge fixed by the use of a Faddeev Popov determinant, thereby removing the unphysical degrees of freedom that are present beforehand. A quadratic GRIN material has been used as an explicit example to see how this works in practice. The Gouy phase and longitudinal field propagation have then been calculated within this framework.

Future investigations could address a number of possibilities. Firstly it would be interesting to further develop the ideas in~\cite{Wan:20} and ~\cite{Wang:21} and make use more fully of the spin coherent states and Haar measure of $SU(2)$ for path integrals of spin. Next, it would be interesting to further explore the classical-quantum interface. For example in~\cite{Qian:11,Qian:18} and \cite{Holleczek:11}, descriptions of how entanglement, duality and cylindrically polarized states of light feature at this interface, and these should be amenable to the current path integral approach (see also~\cite{shen2021} for recent results exploring classical entanglement). Finally, it would be interesting to use the path integral in connection with polarization knots~\cite{Sugic:18,Dalhuisen_2012} and geometric phase~\cite{Dennis_2019}, as an alternative approach. The path integral has been used in polymer and mathematical physics~\cite{kleinert2009path} in conjunction with the theory of knots, so it is likely it would be of use in these types of light fields.

\section*{Acknowledgments}


I thank John Gracey and Andy Wood for useful discussions.

\section*{Disclosures}

The author declares no conflicts of interest.

\bibliography{refs_pol}



\end{document}